%% file: Ong_LiBatt_Rev.tex
\documentclass[
journal = jpccck,
layout = traditional
]{achemso}

\usepackage{color}
\usepackage[normalem]{ulem}
\setkeys{acs}{usetitle = true}

\newcommand{\tit}[1]{\textit{#1}}		
\newcommand{\tsf}[1]{\textsf{#1}}		
\SectionNumbersOn

\newcommand{\reax}[0]{\tsf{ReaxFF}}

\author{Mitchell T. Ong}
\email{ong7@llnl.gov}
\affiliation[Materials Science Division, Lawrence Livermore National Laboratory]
{Materials Science Division, Lawrence Livermore National Laboratory, Livermore, CA 94550, USA}

\author{Osvalds Verners}
\affiliation[Department of Mechanical and Nuclear Engineering, Pennsylvania State University]
{Department of Mechanical and Nuclear Engineering, Pennsylvania State University, University Park, PA 16801, USA}

\author{Erik W. Draeger}
\affiliation[Center for Applied Scientific Computing, Lawrence Livermore National Laboratory]
{Center for Applied Scientific Computing, Lawrence Livermore National Laboratory, Livermore, CA 94550, USA}

\author{Adri C. T. van Duin}
\affiliation[Department of Mechanical and Nuclear Engineering, Pennsylvania State University]
{Department of Mechanical and Nuclear Engineering, Pennsylvania State University, University Park, PA 16801, USA}

\author{Vincenzo Lordi}
\email{lordi2@llnl.gov}
\affiliation[Materials Science Division, Lawrence Livermore National Laboratory]
{Materials Science Division, Lawrence Livermore National Laboratory, Livermore, CA 94550, USA}

\author{John E. Pask}
\affiliation[Physics Division, Lawrence Livermore National Laboratory]
{Physics Division, Lawrence Livermore National Laboratory, Livermore, CA 94550, USA}

\title{Lithium Ion Solvation and Diffusion in Bulk Organic Electrolytes from First Principles and Classical Reactive Molecular Dynamics}


\begin{document}

\begin{tocentry}
\includegraphics{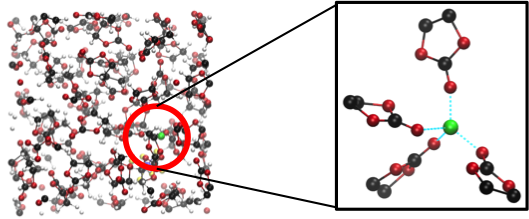}
Favorable solvation structure of Li$^+$ in ethylene carbonate electrolyte.
\end{tocentry}


\begin{abstract}
Lithium-ion battery performance is strongly influenced by the ionic conductivity of the electrolyte, which depends on the speed at which Li ions migrate across the cell and relates to their solvation structure. 
The choice of solvent can greatly impact both solvation and diffusivity of Li ions. 
We use first principles molecular dynamics to examine the solvation and diffusion of Li ions in the bulk organic solvents ethylene carbonate (EC), ethyl methyl carbonate (EMC), and a mixture of EC/EMC. 
We find that Li ions are solvated by either carbonyl or ether oxygen atoms of the solvents and sometimes by the PF$_6^-$ anion. 
Li$^+$ prefers a tetrahedrally-coordinated first solvation shell regardless of which species are involved, with the specific preferred solvation structure dependent on the organic solvent.
In addition, we calculate Li diffusion coefficients in each electrolyte, finding slightly larger diffusivities in the linear carbonate EMC compared to the cyclic carbonate EC.
The magnitude of the diffusion coefficient correlates with the strength of Li$^+$ solvation. 
Corresponding analysis for the PF$_6^-$ anion shows greater diffusivity associated with a weakly-bound, poorly defined first solvation shell.
These results may be used to aid in the design of new electrolytes to improve Li-ion battery performance.
\end{abstract}


\section{Introduction}

There is a growing need to replace gasoline and other fossil fuels with environmentally-friendly alternative energy sources.\cite{GooKim10}
However, many of these alternative energy sources such as solar, wind, waves, and geothermal energy require advances in storage technology in order to become practical.
Li-ion batteries are convenient, portable energy storage devices, which are currently used to power many handheld consumer devices and electric vehicles. 
Main components of a Li-ion battery include the anode, cathode, and electrolyte. 
Carbonaceous materials like graphite are often used for the anode due to the low cost of carbon and the ease with which Li intercalates into the material. 
Lithium transition metal oxides are frequently used for the cathode. 
The electrolyte provides the medium through which Li ions diffuse between the anode and cathode and usually consists of a lithium salt dissolved in either an organic liquid, ionic liquid, or gel polymer.\cite{Xu04}  

The choice of the electrolyte can influence the overall performance of the Li-ion battery.\cite{Xu04}
Many existing batteries use organic liquids to dissolve the Li salt. 
Among the most common organic liquids used in commercial batteries today are various carbonates, including ethylene carbonate (EC), ethyl methyl carbonate (EMC), and mixtures of these. 
LiPF$_6$ is a frequently used salt in the electrolyte as it exhibits high ionic conductivity.
A good organic solvent will be able to dissolve a high concentration of salt, resulting in a high dielectric coefficient. 
A low solvent viscosity facilitates ionic transport.\cite{Xu04} 
Typically, cyclic carbonates like EC have a high dielectric constant, but also have high viscosity, while linear carbonates like EMC have lower viscosity, but also a low dielectric constant. 
Moreover, some organic liquids like EC have a  melting point above room temperature, so that they are not liquids over the entire operating temperature range.
In order to resolve these issues and optimize the viscosity, dielectric constant, and melting point of the electrolyte for battery performance, linear and cyclic carbonates are often mixed.\cite{Xu04} 
EC is a commonly used liquid for mixed electrolytes since it is known to form a protective layer known as the solid electrolyte interphase (SEI)\cite{KimDuiShe11, VatBorSmi12, GanKenJia12, Leu13, NieChaLuc13, JorKumAbr13, OgaOhbKou13} on graphitic anodes, which prevents excessive electrolyte decomposition and promotes reversible intercalation into and out of the anode. 

Experiments have been conducted to examine the effect of the electrolyte composition on Li transport and solvation using different spectroscopic techniques such as FTIR,\cite{BarBucWis00} Raman,\cite{CazMusBen96, MorAsaYos98, KlaAroNaz98, DouRevLau99} nuclear-magnetic resonance (NMR),\cite{CazMusBen96, YanXiaLuc10, BogVazGre13} electrospray ionization mass spectroscopy,\cite{FukMatHas01} neutron scattering,\cite{KamYasTak07} and X-ray diffraction.\cite{SoeMilMai01}  
Some of these experiments attempted to determine the coordination number around the Li$^+$ in using different Li salts, solvents, and concentrations.\cite{CazMusBen96, MorAsaYos98, KamYasTak07} 
Reported coordination numbers in these works range from $\sim$2 to 5. 
However, there has been relatively little experimental characterization of the solvent molecule orientation around Li$^+$.
One such study was performed by Cazzanelli \tit{et al.}, who determined a coordination number of $\sim$2 in a mixture of EC and propylene carbonate (PC) at different concentrations.\cite{CazMusBen96}
At high concentrations, they concluded that Li$^+$ was ``sandwiched'' between two ring solvent molecules.
Recent NMR experiments have also shown that there is a preference for EC to solvate Li$^+$ over DMC in mixed EC/DMC systems and that the carbonyl oxygen atoms are involved in solvation.\cite{BogVazGre13} 
Separate NMR experiments on transport properties have also been carried out, in which experimental diffusion coefficients for Li$^+$ in different electrolytes were found to range from $1-8\times10^{-6}$~cm$^2$/s at 30$^{\circ}$C.\cite{HayAihAra99}  
Mixed EC/EMC systems have Li$^+$ diffusion coefficient determined to be between $1.5-4.5 \times 10^{-6}$~cm$^2$/s, depending on salt concentration.\cite{CapSaiKag99} 

Theoretical work has also been performed to understand Li transport and solvation in various carbonate electrolytes. 
To date, many of these simulations have been performed using classical force fields\cite{BorSmi06, BorSmi09, SoeMilMai98, LiBal99, Tas02, MasProRey04, TasGolWin11}. 
Some of these studies indicate a coordination number of 4 where the Li$^+$ interacts with carbonyl oxygen atoms of the carbonate.\cite{BorSmi06, BorSmi09, SoeMilMai98, MasProRey04, TasGolWin11} 
In mixed carbonate systems such as EC/DMC, it was found that both EC and DMC participate in solvating Li$^+$.\cite{BorSmi06} 
Furthermore, there was a greater affinity for Li$^+$ to dissociate from its counter-ion in cyclic carbonates relative to linear carbonates.\cite{BorSmi06} 
However, classical potentials are limited in their transferability and their ability to describe charge transfer effects.
These limitations are not present in first-principles methods, which treat the electrons quantum mechanically. 
Static quantum calculations using cluster models have been used to study the energetics of different solvation structures,\cite{YuBalBud11,WanNakUe01,WanBal05} but these studies do not include the effect of the overall liquid environment. 

Only recently have first principles molecular dynamics (FPMD) based on forces from density functional theory been used to study the solvation and transport properties of Li$^+$ in different electrolytes.\cite{LeuBud10, GanJiaKen11, BhaChoCho12}
Use of FPMD can be more predictive than classical force fields due to its parameter-free nature.
FPMD has better transferability and can more accurately describe polarizability, charge transfer, and partial charges than classical potentials.
It also provides a better basis for future comparative studies of the electrolyte solutions near interfaces and in reactive environments.
Previous work using FPMD by both Leung \tit{et al.}\cite{LeuBud10} and Ganesh \tit{et al.}\cite{GanJiaKen11} found Li$^+$ solvation structures that agreed with previous classical force field simulations and were generally consistent with experiments.
In addition, Ganesh calculated diffusion coefficients that were slightly higher than experimental values and previous theoretical work.\cite{GanJiaKen11}
However, both of these works considered systems with no more than $\sim$300--400 atoms and simulation times of 13--25~ps, which may have limited the ability to systematically extract quantities of interest.
Both also used certain approximations, such as increased hydrogen masses and large time steps, to make the simulations more computationally feasible. 
Therefore, it remains unclear how accurately these simulations fully describe Li$^+$ solvation and diffusion in real systems.

In this work, we carry out FPMD simulations to study solvation and diffusion in several experimentally relevant carbonate-based organic Li-ion battery electrolytes.
We compare the cyclic carbonate solvent EC to the linear carbonate EMC, as well as an EC/EMC mixture. The typical LiPF$_6$ salt is chosen for this work.
Solvation and transport properties for the EC/LiPF$_6$ system have been previously studied with FPMD,\cite{LeuBud10, GanJiaKen11,BhaChoCho12} but similar linear carbonate and cyclic/linear carbonate mixtures have been explored only using classical force fields.\cite{BorSmi09}
We analyze solvation structures of Li$^+$  to examine how the choice of solvent influences the structure and explore the range of possible solvation structures in each solvent.
Furthermore, we examine the interaction of the Li$^+$ and PF$_6^-$ and compare their solvation properties.
We calculate the diffusion coefficient in the different electrolytes to understand why Li$^+$ diffuses faster in one solvent than another and find correlations between solvation and diffusivity.
The relation of solvation and diffusivity of PF$_6$ is also studied and compared to that of Li$^+$.
Finally, in addition to performing these studies with larger system sizes and longer time scales than previous FPMD simulations, we further employ the \reax{} force field\cite{reaxff} to quantify the effects of finite size and time scales on the observed Li$^+$ solvation structures and diffusivity.
We anticipate that our findings can be used to design new electrolytes that will improve the cycling rate in batteries by tuning solvation to enhance diffusivity.

\section{Computational Details}
\label{sec:CompDetails}

We perform first principles molecular dynamics using density functional theory (DFT) with the projector augmented wave (PAW) method\cite{Blo94, KreJou99} and the PBE generalized gradient approximation exchange-correlation functional\cite{PerBurErn96,PerBurErn97}, as implemented in the \tsf{VASP}\cite{vasp, KreFur96} software package. 
A 450 eV plane-wave cutoff was used with Brillouin zone sampling restricted to the $\Gamma$ point.
All molecular dynamics simulations were performed in the NVT ensemble using a Nos\'e-Hoover thermostat\cite{Nos84, Hoo85}, with the Nos\'e frequency of $\sim$1000~cm$^{-1}$ corresponding to a period of $\sim$32~fs, and time step of 0.5~fs. 
Each system was equilibrated for 5--7.5~ps at 330~K, followed by 30~ps of simulation time  to gather statistics.
A temperature of 330~K was used to mimic an intermediate Li-ion battery operating temperature and to ensure that EC was not frozen ($T_\textrm{melt}^\textrm{EC} = 310$~K).

Solvation structures were characterized with pair correlation functions, calculated using a bin size of 0.03~\AA.
Average coordination numbers were computed from the integral of the pair correlation function.
We further quantify how tightly the ions are solvated by calculating the average residence time of first shell solvent molecules by fitting an exponential to the time correlation function
\begin{equation} 
\label{eq:eqn1}
P_\textrm{solv} (t)=\langle H(t) \cdot H(0) \rangle,
\end{equation}
where $H(t)$ is 1 if a given molecule is within the first solvation shell and 0 otherwise.\cite{ImpMadMcd83, BorSmi06, BorSmi09, WooMar07}
The distance cutoff of the first solvation shell is taken from the first minimum in the pair correlation function between the solvated ion and the center of mass of each solvent molecule.
A stretched exponential of the form $\exp [ -(t/\tau)^\beta ]$ gives the best fits, with $\tau$ being the residence time and $\beta$ an adjustable parameter.\cite{BorSmi06, BorSmi09}

Diffusion coefficients for Li$^+$ and PF$_6^-$ were extracted using two methods: (i) integration of the velocity autocorrelation function (VACF) via \cite{Gre54, Kub57},
\begin{equation} 
\label{eq:eqn2}
D=\frac{1}{3} \int\limits_0^\infty \frac{1}{N} \sum_{i=1}^{N} \langle v_i (0) \cdot v_i(t)\rangle dt, 
\end{equation}
and (ii) analysis of the mean square displacement (MSD) over time using the Stokes-Einstein relation,\cite{Ein05} 
\begin{equation}
\label{eq:eqn3}
D = \frac{1}{6} \frac{\langle (\delta r)^2 \rangle}{\Delta t}. 
\end{equation}
For infinite statistics, using either the MSD or VACF to calculate the diffusion coefficient should produce the exact same answer as they are mathematically equivalent.
However, for finite statistics, they are not numerically identical,\cite{Woo01} so we calculate the diffusion coefficient using both methods.
The VACF and MSD were both calculated by averaging over multiple trajectory windows spanning the entire trajectory with starting configurations every 50~fs, using various window lengths from 5-15~ps in increments of 2.5~ps.
Equation~\ref{eq:eqn2} was used to calculate $D$ from the VACF for each window length.
Likewise, the slope of the linear regime in the MSD was used to calculate $D$ from Eq.~\ref{eq:eqn3} for each window length.
In each case, the values for $D$ for each window length were averaged to get a final estimate of the diffusion coefficient.
Reported uncertainties reflect the standard deviation in this average. 

\section{Results and Discussion}

\subsection{Li$^+$ Solvation in Ethylene Carbonate}

\begin{figure}
   \centering
   \includegraphics[width=0.75\textwidth,clip=true]{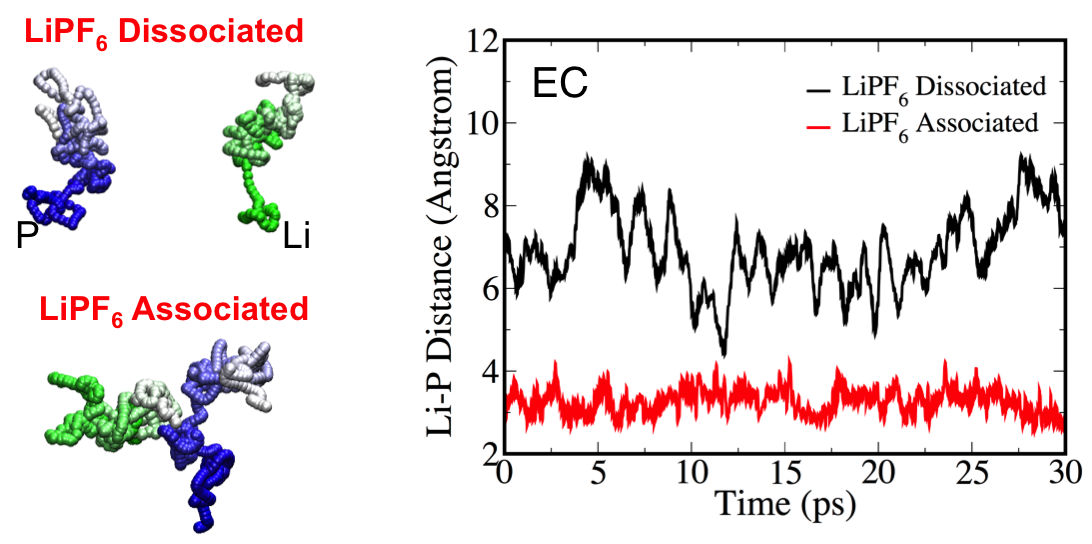}
   \caption{(Left) Trajectories for Li$^+$ and P atom of PF$_6^-$ for 63 EC + 1 LiPF$_6$ system. Color gradients designate time, where darker colors are earlier in time. (Right) Li--P distance as a function of time for trajectories that are initially dissociated or associated.}
   \label{fig:ECIonMotion}
\end{figure}

We performed first principles molecular dynamics simulations of a single LiPF$_6$ molecule dissolved in a periodic box of 63 EC molecules. 
This system corresponds to 638 total atoms, with a density 1.32~g/cc and Li concentration of 0.23~M. 
Typical Li concentration in  commercial batteries is 4-5 times larger, but this concentration was used to focus on the solvation of a single Li$^+$ without the effects of other salt molecules being present.
We carried out two independent simulations where the Li$^+$ and PF$_6^-$ ions were initially either associated or dissociated. 
In Fig.~\ref{fig:ECIonMotion}, we show the Li--P distance over each of the two trajectories, which indicate that the LiPF$_6$ remains either associated or dissociated for the entire simulation. 
In addition, we display the trajectories of the Li$^+$ ion and the P from the PF$_6^-$ ion where the color gradients (dark to light) indicate time.
We find that Li$^+$ and PF$_6^-$ follow similar trajectories even when dissociated, with separation $\sim$5--8~\AA, giving evidence of correlated motion.

\begin{figure}
   \centering
   \includegraphics[width=1.00\textwidth,clip=true]{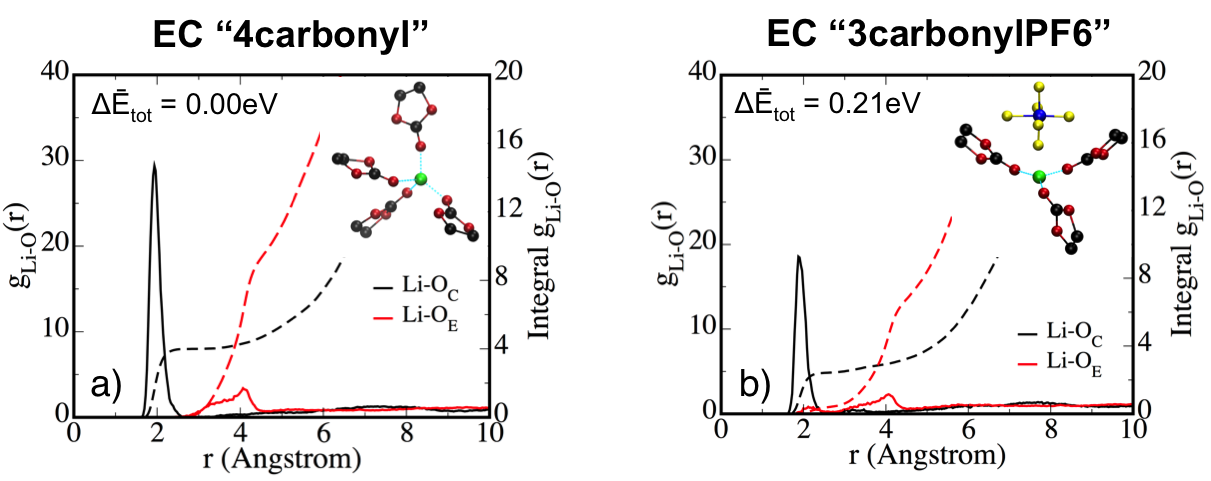}
   \caption{Li--O$_C$ and Li--O$_E$ pair correlation functions (solid lines) and their integrals (dashed lines) for (a) ``4carbonyl'' and (b) ``3carbonylPF6'' solvation structures of EC. Snapshot of solvation structure and average energy relative to the lowest energy structure are shown in the inset.}
   \label{fig:ECSolvation}
\end{figure}

\begin{figure}
   \centering
   \includegraphics[width=1.00\textwidth,clip=true]{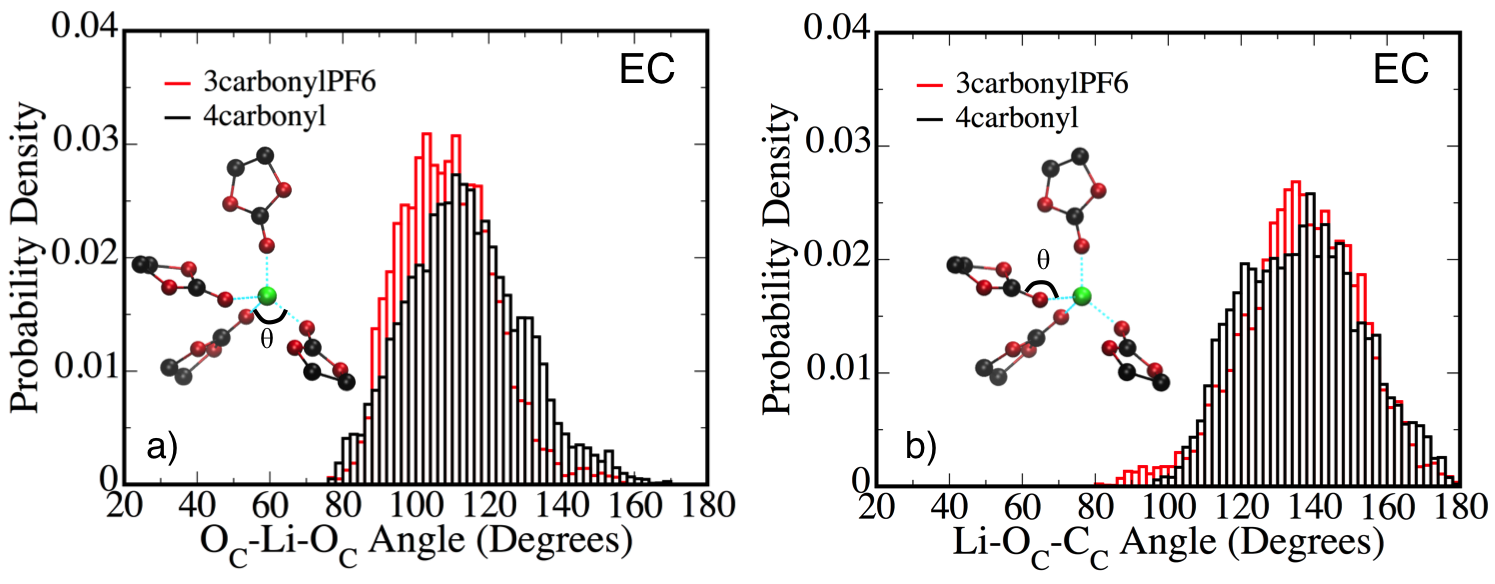}
   \caption{Histogram of a representative (a) O$_C$--Li--O$_C$ angle and (b) Li--O$_C$--C$_C$ angle (denoted in the insets) for the ``3carbonylPF6'' and ``4carbonyl'' solvation structures of EC during the trajectory.}
   \label{fig:ECOrientation}
\end{figure}

The pair correlation functions between the Li ion and either the carbonyl oxygen atoms from EC, designated O$_C$, or the ether oxygen atoms, designated O$_E$, are shown in Fig.~\ref{fig:ECSolvation}. 
We find two different solvation structures, one where the PF$_6^-$ stays apart from Li$^+$ [Fig.~\ref{fig:ECSolvation}(a)] and another where they remain close [Fig.~\ref{fig:ECSolvation}(b)]. 
For the case where they remain associated, we observe an average coordination number of $\sim$2.5 for Li--O$_C$ and $\sim$0.5 for Li--O$_E$ in the first solvation shell. 
In this case, the PF$_6$ also occupies a site in the first solvation shell with a total coordination number of 4.
We label this solvation structure ``3carbonylPF6.''
In the case where Li and PF$_6$ are dissociated, we calculate a coordination number of 4 for Li--O$_C$.
We denote this solvation structure ``4carbonyl.''
Representative snapshots of the ``3carbonylPF6'' and ``4carbonyl'' solvation structures are shown in Fig.~\ref{fig:ECSolvation}. 
The peak of the Li--O$_C$ pair correlation function at approximately $\sim$1.9~{\AA} agrees with previous classical\cite{SoeMilMai98, BorSmi09, LiBal99} and FPMD\cite{GanJiaKen11, BhaChoCho12} simulations which range between 1.7--2.0~\AA. Our total coordination number of 4 also agrees with previous theoretical work\cite{BorSmi06, SoeMilMai98, MasProRey04, TasGolWin11, LeuBud10, GanJiaKen11, BhaChoCho12} and experiments.\cite{KamYasTak07}
We compare the thermodynamic stability of these two structures by computing the average relative energies over the trajectories, which are also indicated in Fig.~\ref{fig:ECSolvation}.
We find the ``4carbonyl'' solvation structure to be favorable by $\sim$0.2~eV.
We also examined the orientation of the solvent molecules around Li$^+$ in each case by tracking the  O$_C$--Li-O$_C$ angle. 
Fig.~\ref{fig:ECOrientation}(a) shows the histogram of the O$_C$--Li-O$_C$ angles for both the ``3carbonylPF6'' and ``4carbonyl'' solvation structures. 
We see that both are peaked at  $\sim$110$^{\circ}$, indicating a preference for a tetrahedral arrangement in both cases. 
We also find that the carbonyl group of the EC molecule tends to point toward the Li$^+$, with Li--O$_C$--C$_C$ angle $\sim$140$^\circ$ as shown in Fig.~\ref{fig:ECOrientation}(b).
Previous FPMD simulations by Ganesh\cite{GanJiaKen11} have also determined this angle to be 140$^{\circ}$, which is consistent with the experimental value of 138$^{\circ}$,\cite{KamYasTak07} but lower than the classical force field value\cite{BorSmi09} of 150$^{\circ}$.
The tetrahedral pattern we observe for ``4carbonyl'' agrees with previously calculated solvation structures for Li$^+$ in EC.\cite{BorSmi06, SoeMilMai98, MasProRey04, TasGolWin11, LeuBud10, GanJiaKen11, BhaChoCho12}
The near 110$^{\circ}$ angle for ``3carbonylPF6'' suggests that regardless of the composition of the first solvation shell, Li$^+$ prefers to be solvated in a tetrahedral fashion.

\subsection{Li$^+$ Solvation in Ethyl Methyl Carbonate}

\begin{figure}
   \centering
   \includegraphics[width=0.75\textwidth,clip=true]{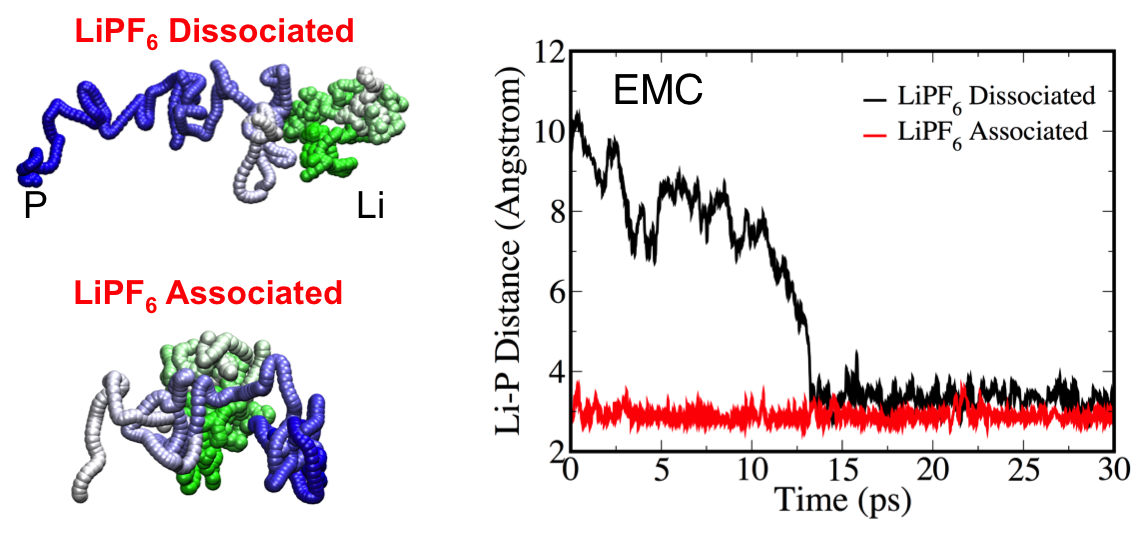}
   \caption{(Left) Trajectories for Li$^+$ and P atom of PF$_6^-$ for 42 EMC + 1 LiPF$_6$ system. Color gradients designate time, where darker colors are earlier in time. (Right) Li--P distance as a function of time for trajectories that are initially dissociated or associated.}
   \label{fig:EMCIonMotion}
\end{figure}

Unlike EC, ethyl methyl carbonate (EMC) has a lower viscosity and freezes at a much lower temperature, but it also has a lower dielectric constant. 
We examined how the solvation structures and ionic motion differ when a linear carbonate such as EMC is used as the solvent. 
The simulation system consisted of 42 EMC molecules and 1 LiPF$_6$ (638 total atoms) with a density of 1.01~g/cc and concentration of 0.22~M.
Again, we ran simulations with the LiPF$_6$ initially either associated or dissociated. 
We found that the initially dissociated LiPF$_6$ re-associated within 15~ps during the simulation, as shown in the Li--P distance plot in Fig.~\ref{fig:EMCIonMotion}.
On the other hand, when LiPF$_6$ started associated, it remained associated throughout the simulation. 
Upon examination of their trajectories, we found that when Li$^+$ and PF$_6^-$ are initially dissociated, the PF$_6^-$ moves toward and finds the Li$^+$.
Overall, PF$_6^-$ is observed to migrate further than Li$^+$.
This is likely due to the Li$^+$ being more tightly solvated than the PF$_6^-$, as discussed later.

\begin{figure}
   \centering
   \includegraphics[width=1.00\textwidth,clip=true]{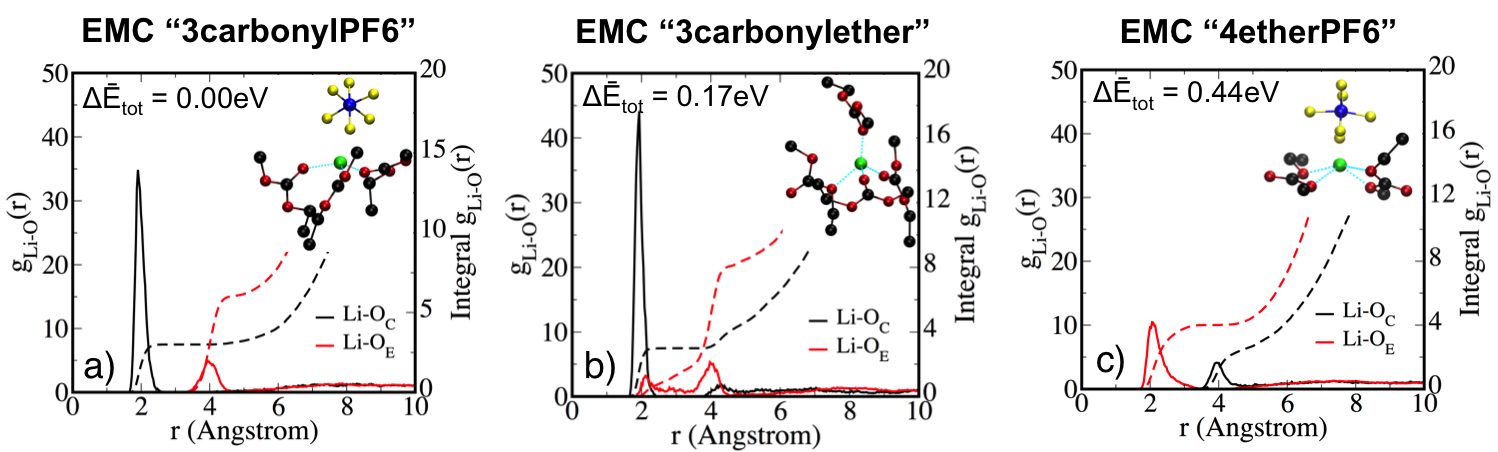}
   \caption{Li--O$_C$ and Li--O$_E$ pair correlation functions (solid lines) and their integrals (dashed lines) for (a) ``3carbonylPF6,'' (b) ``3carbonylether,'' and (c) ``4etherPF6'' solvation structure of EMC. Snapshot of solvation structure and average energy relative to the lowest energy structure are shown in the inset.
   }
   \label{fig:EMCSolvation}
\end{figure}

\begin{figure}
   \centering
   \includegraphics[width=1.00\textwidth,clip=true]{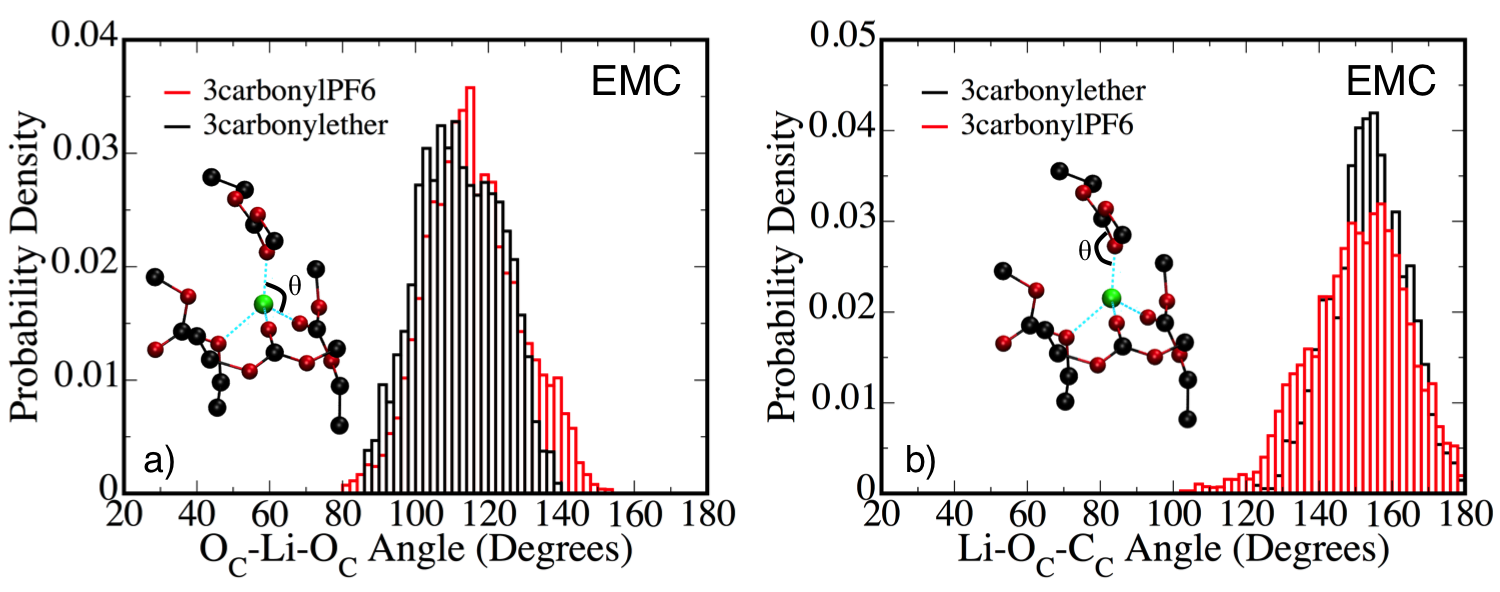}
   \caption{Histogram of a representative (a) O$_C$--Li--O$_C$ angle and (b) Li--O$_C$--C$_C$ angle (denoted in the insets) for ``3carbonylPF6'' and ``3carbonylether'' solvation structures of EMC during the trajectory.}
   \label{fig:EMCOrientation}
\end{figure}

We observe three different solvation structures during the simulations. 
The pair correlation functions of Li--O$_C$ and Li--O$_E$ for each of these solvation structures along with representative snapshots and relative energies are shown in Fig.~\ref{fig:EMCSolvation}. 
We find that the lowest energy structure is one where the Li$^+$ and PF$_6^-$ are associated [Fig.~\ref{fig:EMCSolvation}(a)].
This is a significant difference compared to EC.
The total coordination number of Li$^+$ in this preferred structure is 4, consisting of 3 carbonyl oxygens from 3 EMC molecules and the nearby PF$_6^-$.
We designate this solvation structure as ``3carbonylPF6."
We note that a similar structure was also observed with EC, but it is not the lowest energy configuration.
There is another solvation structure in EMC that is 4-fold coordinated, and this consists of 4 EMC molecules oriented such that Li$^+$ is solvated by 3 carbonyl oxygen atoms and one ether oxygen atom.
In this case, the Li$^+$ and PF$_6^-$ remain apart; we denote this structure as ``3carbonylether" [Fig.~\ref{fig:EMCSolvation}(b)].
The orientations of these two 4-fold coordinated solvation structures are analyzed in Fig.~\ref{fig:EMCOrientation}(a), where we plot the O$_C$--Li--O$_C$ bond angle for the ``3carbonylPF6'' and ``3carbonylether'' cases.
For both these cases, there is a peak near 110$^{\circ}$, similar to EC, indicating a preferred tetrahedral arrangement of the solvent molecules regardless of whether EMC or PF$_6^-$ is solvating Li$^+$. 
We also see a preference for the carbonyl oxygen to point toward Li$^+$, from the Li--O$_C$--C$_C$ angle of $\sim$150$^\circ$ shown in Fig.~\ref{fig:EMCOrientation}(b). 
Similar conclusions were drawn for EC, but the Li--O$_C$--C$_C$ angle is slightly bigger for EMC than EC, which is also consistent with trends seen by Borodin \tit{et al.},\cite{BorSmi09} using classical force fields.
Unlike for EC, however, for EMC we also found a non-tetrahedral structure where the Li$^+$ is solvated by 4 total ether oxygen atoms belonging to two EMC molecules (2 ether oxygen atoms per EMC) and PF$_6^-$, in a square pyramidal-like fashion.
We label this structure ``4etherPF6'' [Fig.~\ref{fig:EMCSolvation}(c)].
This structure is not energetically preferred, being more than 0.4~eV higher in energy than ``3carbonylPF6.'' 
For EMC, a corresponding ``4carbonyl'' solvation structure, which is favored by EC, is not observed in either of the trajectories.
Steric issues likely prohibit this structure for EMC, since the length of the molecule makes it unfavorable to have four EMC molecules around Li$^+$. 
With PF$_6^-$ similar in size to EC, the ``3carbonylPF6'' structure is preferred for EMC, instead of the ``4carbonyl'' structure as for EC.

\subsection{Li$^+$ Solvation in 3:7 Mixture of EC/EMC}

\begin{figure}
   \centering
   \includegraphics[width=0.75\textwidth,clip=true]{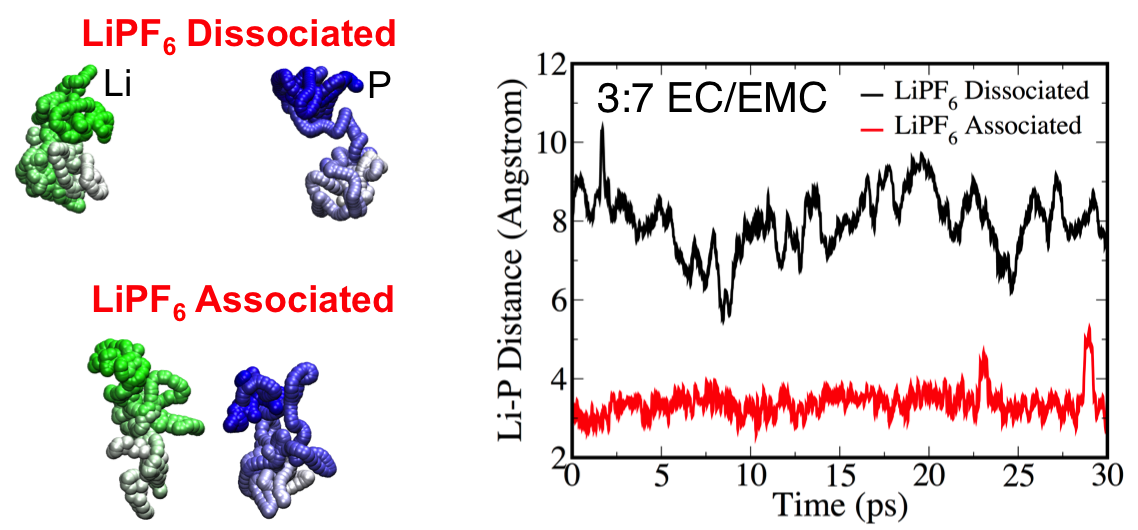}
   \caption{(Left) Trajectories for Li$^+$ and P atom of PF$_6^-$ for 15 EC + 35 EMC + 1 LiPF$_6$ system. Color gradients designate time, where darker colors are earlier in time.  (Right) Li--P distance as a function of time for trajectories that are initially dissociated or associated.}
   \label{fig:ECEMCIonMotion}
\end{figure}

We also examined the effect of mixing different organic solvents for the electrolyte. 
We studied a mixture of EC and EMC in a 3:7 ratio, mimicking previous experiments.\cite{NieChaLuc13} 
The simulation system consisted of 15 EC molecules, 35 EMC molecules, and 1 LiPF$_6$ (683 total atoms) with a density of 1.165~g/cc and a concentration of 0.23~M.
As in the previous studies, we began the simulations with the LiPF$_6$ either associated or dissociated. 
In Fig.~\ref{fig:ECEMCIonMotion}, we plot the Li--P distance for both cases and show the trajectories of the two ions.
We see that the ions remain associated or dissociated for the duration of the simulation. 
In addition, based on the trajectories, we see that the Li and P in both the associated and dissociated cases follow very similar paths, indicative 
 that the ions behave more like in EC than in EMC, with a large degree of correlated motion even when the ions are separated.
 This similarity to EC occurs even with the mixture containing only 30\% EC.
Therefore, adding only a small percentage of EC to the system results in a dramatic change in the ion motion with respect to one another.

\begin{figure}
   \centering
   \includegraphics[width=1.00\textwidth,clip=true]{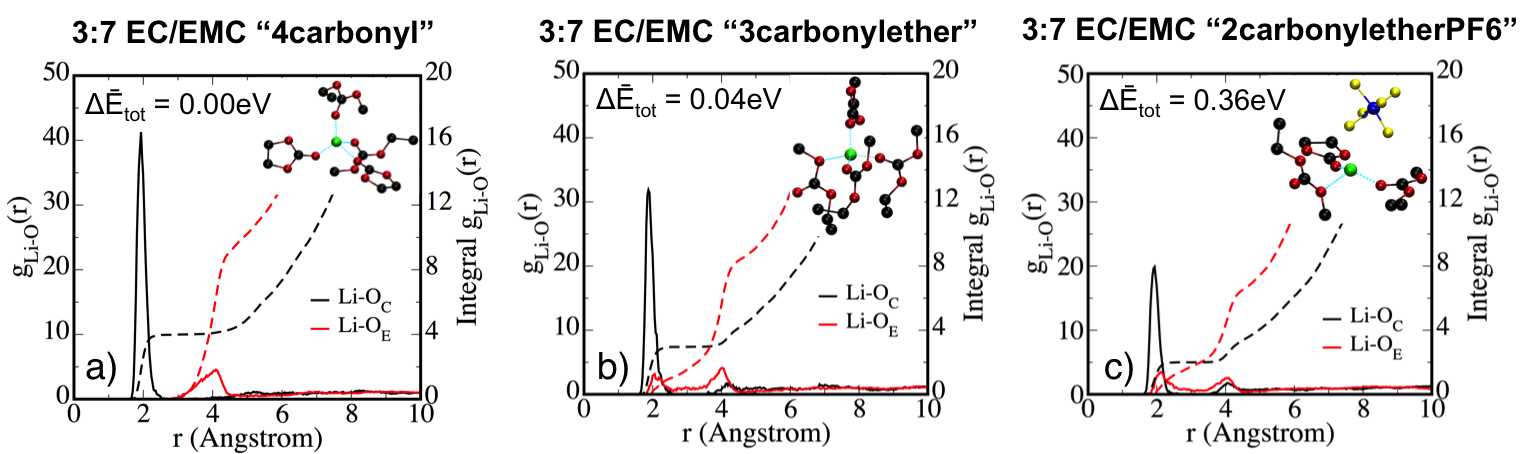}
   \caption{Li--O$_C$ and Li--O$_E$ pair correlation functions (solid lines) and their integrals (dashed lines) for (a) ``4carbonyl,'' (b) ``3carbonylether,'' and (c) ``2carbonyletherPF6'' solvation structures of mixed EC/EMC. Snapshot of solvation structure and average energy relative to the lowest energy structure are shown.
   }
   \label{fig:ECEMCSolvation}
\end{figure}

From these trajectories, we find three different solvation structures:  two structures where the Li$^+$ and PF$_6^-$ ions stay apart and one structure where they remain together. 
Snapshots of each solvation structure, their pair correlation functions for Li--O$_C$ and Li--O$_E$, and their relative energies are shown in Fig.~\ref{fig:ECEMCSolvation}. 
For the two solvation structures with LiPF$_6$ dissociated, one shows a coordination number of 4 for Li--O$_C$, indicating that 4 carbonyl oxygen atoms  solvate Li$^+$;
the other has a coordination number of $\sim$3 for Li--O$_C$ and  $\sim$1 for Li--O$_E$, where 3 carbonyl oxygen atoms and 1 ether oxygen atom solvate Li$^+$. 
We denote these solvation structures ``4carbonyl'' [Fig.~\ref{fig:ECEMCSolvation}(a)] and ``3carbonylether'' [Fig.~\ref{fig:ECEMCSolvation}(b)], respectively. 
For the one structure where LiPF$_6$ is associated, there is a coordination number of 2 for Li--O$_C$ and  $\sim$1 for Li--O$_E$, indicating 2 carbonyl oxygen atoms and 1 ether oxygen atom solvate Li$^+$ along with the PF$_6^-$. 
We label this structure ``2carbonyletherPF6'' [Fig.~\ref{fig:ECEMCSolvation}(c)].
The ``4carbonyl''  and ``3carbonylether'' solvation structures are nearly energetically equivalent with only a 0.04~eV difference between the two,
whereas the ``2carbonyletherPF6'' is more than 0.3~eV higher in energy.
In this 3:7 EC/EMC mixture, Li$^+$ prefers to be separated from PF$_6^-$, similar to pure EC.
In addition, we observe that one EMC in the ``3carbonylether'' solvation structure is replaced by an EC molecule during one of the trajectories, leaving 2 EC molecules and 2 EMC molecules involved in the first solvation shell and forming the ``4carbonyl'' solvation structure.
This result indicates a strong preference for EC molecules to solvate Li$^+$, considering the small fraction of EC in the system.

\begin{figure}
   \centering
   \includegraphics[width=1.00\textwidth,clip=true]{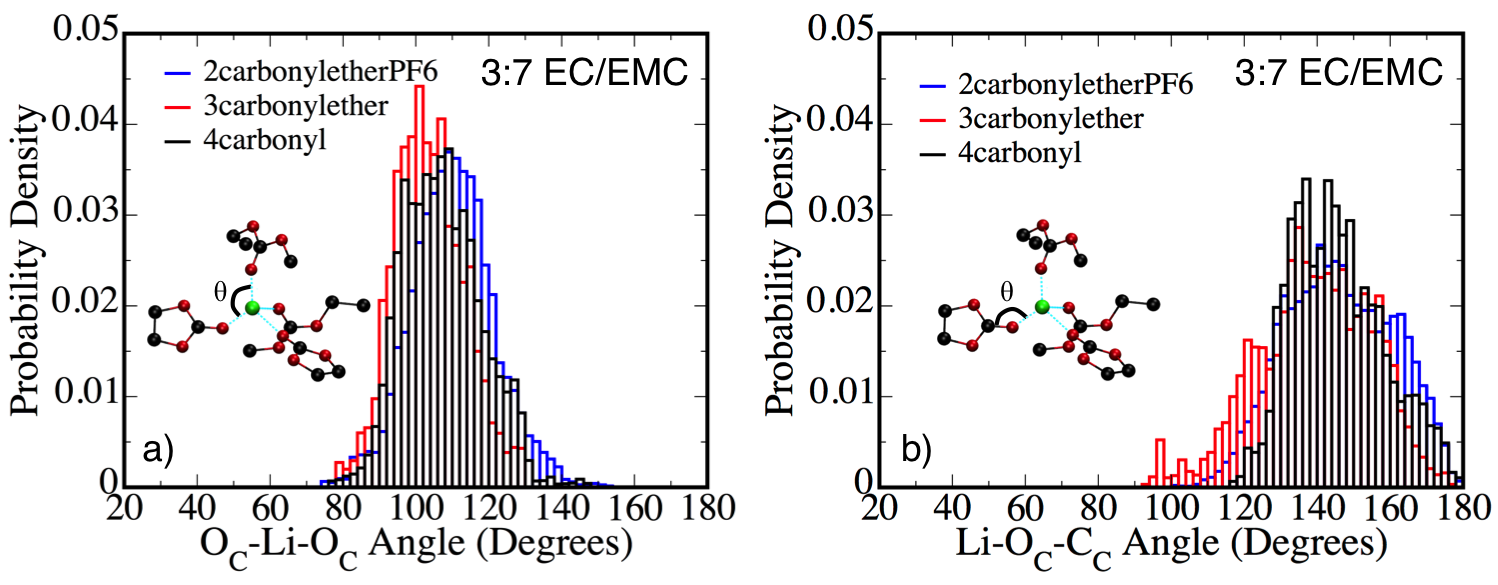}
   \caption{Histogram of a representative (a) O$_C$--Li--O$_C$ angle and (b) Li--O$_C$--C$_C$ angle (denoted in the insets) for the ``4carbonyl,'' ``2carbonyletherPF6,'' and ``3carbonylether''  solvation structures of mixed EC/EMC.}
   \label{fig:ECEMCOrientation}
\end{figure}

Recent $^{17}$O NMR experiments performed on mixed EC/DMC systems have also shown a strong preference for EC to solvate Li$^+$ as opposed to a linear carbonate such as DMC (or EMC).\cite{BogVazGre13}
This is consistent with our observation that one EMC molecule is replaced by an EC molecule in the first solvation during the course of our simulation.
In addition, classical simulations also observe an equal amount of cyclic carbonate, EC, and linear carbonate, DMC, when the ions are dissociated.\cite{BorSmi09}
Therefore, our results are consistent with experiments and classical simulations and provide strong evidence that Li$^+$ prefers to be solvated by EC when present in these systems.
As described above, this tendency likely is related to steric effects and the fact that EC and PF$_6$ have similar sizes.
We also examined the orientation of the solvent molecules for all three cases. 
In Fig.~\ref{fig:ECEMCOrientation}(a), we show histograms of the O$_C$--Li--O$_C$ angle for each solvation structure. 
Again, we find  the peak of the histogram for all three cases near 110$^{\circ}$, with Li$^+$ preferring  a tetrahedral solvation structure. 
In Fig.~\ref{fig:ECEMCOrientation}(b), we plot the Li--O$_C$--C$_C$ angle, and find that the carbonyl oxygen atoms prefer to point toward Li$^+$ with an angle of $\sim$140$^\circ$, similar to the previous cases (and particularly similar to pure EC).

Overall, our results show that even small variations in the organic solvent can dramatically change the preferred solvation structure, although Li always prefers to be coordinated tetrahedrally in its first solvation shell regardless of which species are around it.

\subsection{Solvation Structure of PF$_6^-$}

\begin{figure}
   \centering
   \includegraphics[width=0.75\textwidth,clip=true]{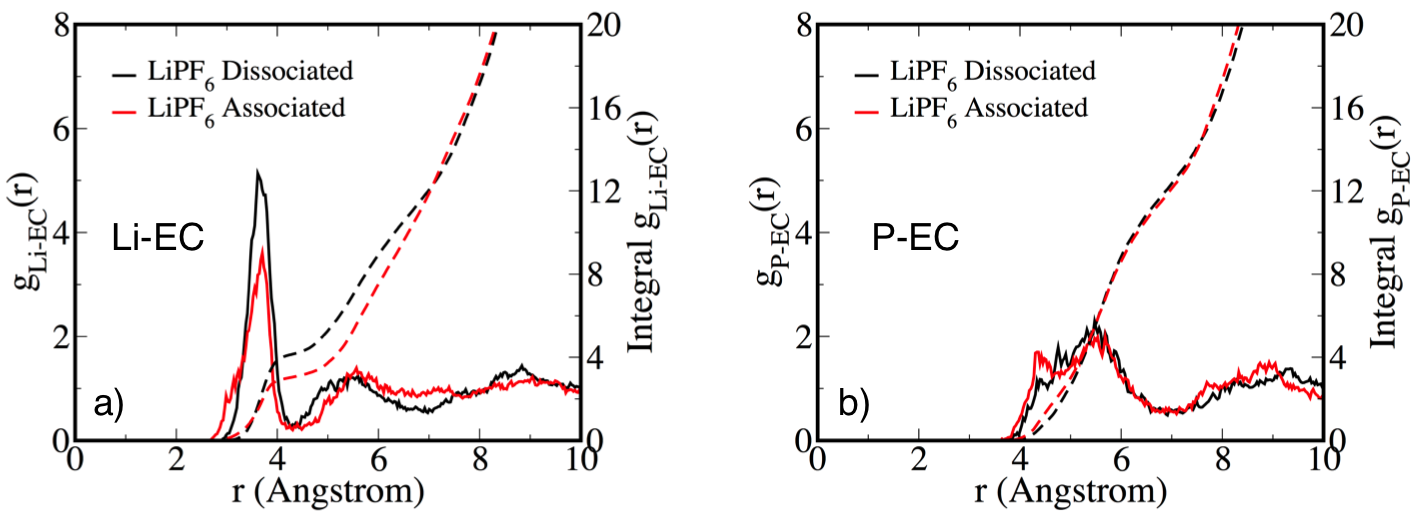}
   \caption{(a) Li--EC and (b) P--EC pair correlation functions (solid lines) and their integrals (dashed lines) for LiPF$_6$ in EC, with the salt molecule initially either associated or dissociated.}
   \label{fig:PF6Solvation}
\end{figure}

Thus far, we have primarily examined the solvation of the Li ion.
We now turn our attention to the solvation structure of the PF$_6^-$ anion and compare it to that of Li$^+$.
There is relatively little discussion about the solvation structure of the counter-ion in previous FPMD and classical simulations in the dissociative limit.
The EC system is used here for illustration, although EMC and the mixture show similar phenomena. 
In Fig.~\ref{fig:PF6Solvation}, we show the pair correlation functions of the center of mass of the EC molecules and either Li [Fig.~\ref{fig:PF6Solvation}(a)] or P [Fig.~\ref{fig:PF6Solvation}(b)], for both trajectories where the ions were initially either associated or dissociated.
The Li--EC pair correlation functions are well structured with a sharp first-shell peak, indicating that Li$^+$ has a well-defined solvation structure with a clear coordination number $\sim$4 in the first solvation shell, as already discussed in detail.
Conversely, the P--EC pair correlation functions are very broad, suggesting that many EC (solvent) molecules dynamically rotate in and out of the first solvation shell on a short time scale.
To quantify the strength of solvation, we computed the average residence time of the first-shell solvent molecules around each of the ions.
For PF$_6^-$, the residence time of EC was 43--90~ps and of EMC was 24--29~ps. (The ranges correspond to variations for trajectories with LiPF$_6$ either associated or dissociated.) 
For Li$^+$, the residence times were well beyond the length of the simulations for both EC and EMC, with the first solvation shells showing little solvent exchange during the trajectories. 
While previous FPMD simulations\cite{GanJiaKen11,BhaChoCho12} for EC report no solvent exchange during their trajectories, we see occasional solvent exchange that preserves the tetrahedral solvation structure.
Thus, we determine that PF$_6^-$ is much more weakly solvated, with a poorly-structured solvation shell and very short solvent molecule residence times, than Li$^+$.
Also, we find that EC solvates PF$_6^-$ somewhat better than EMC, based on the relative residence times.

We have already shown in Figs.~\ref{fig:ECIonMotion}, \ref{fig:EMCIonMotion}, and \ref{fig:ECEMCIonMotion} that PF$_6^-$ appears to move farther than Li$^+$, despite the much heavier mass of the anion.
The increased diffusivity of PF$_6^-$ compared to Li$^+$ is connected to the respective solvation structures and is discussed in detail in the following section.

\subsection{Li$^+$ and PF$_6^-$ Transport Properties}

The transport properties of  Li$^+$ and PF$_6^-$ in each of the electrolytes were studied by comparing their diffusion coefficients. 
Table~\ref{tab:LiDiffusion} compares the Li$^+$ diffusion coefficients for each electrolyte, calculated using both the MSD and VACF methods (see Sec.~\ref{sec:CompDetails} for details). 
We focus on the cases with the solvation in the preferred configurations, which generally corresponds to LiPF$_6$ being dissociated, as discussed above. 
Trends in the diffusion coefficients are generally consistent across the different systems regardless if MSD or VACF is used for the calculation.
Statistical errors of the calculated values are on the order of $1-2 \times 10^{-6}$ cm$^2$/s as shown in Table~\ref{tab:LiDiffusion} or as much as 50\%. 
Improving these errors is nontrivial as it would require running many (>20) independent simulations with one Li ion or a single simulation with many more Li ions (and correspondingly larger system size). 

Values for EC are approximately a factor of two different than previous FPMD results\cite{GanJiaKen11,BhaChoCho12} which calculated the diffusion coefficient to be $\sim1 \times 10^{-5}$ cm$^2$/s at temperatures between 310--450K, likely due to the large uncertainty associated with the short DFT runs.
The highest Li diffusion is seen in EMC.
This is in agreement with experiments\cite{HayAihAra99} and classical simulations\cite{BorSmi09} that find faster diffusion in linear carbonates compared to cyclic carbonates.
Surprisingly, the mixed electrolyte shows slower Li$^+$ diffusion than pure EC, whereas it would be expected to fall between EC and EMC. 
We suspect the reason why the diffusion coefficient of the mixture does not fall between EC and EMC is related to statistical error from the rather short DFT simulation and the variation of solvation structure during the trajectory.
However, we note that the similar solvation structures of Li$^+$ in EC and the EC/EMC mixture result in similar diffusivities in these cases, which are distinct from that in pure EMC.
In fact, the first solvation shell of Li$^+$ in the EC/EMC mixture can contain up to 2 EC molecules, as discussed above so it is expected that the value of the diffusion coefficient would be closer to that of pure EC.

\begin{table}
  \centering
  \begin{tabular}{c c c c c}
    \hline \hline
   & \multicolumn{2}{c}{Li$^+$ Diffusion Coefficient}  \\
   & \multicolumn{2}{c}{(10$^{-6}$~cm$^2$/s)}  \\
  \cline{2-3}
      Electrolyte composition & MSD & VACF \\
    \hline
    63 EC + 1 LiPF$_6$ & 5.2 $\pm$ 0.8 & 7.9 $\pm$ 1.3  \\
    42 EMC + 1 LiPF$_6$ & 9.6 $\pm$ 1.6 & 10.1 $\pm$ 2.1 \\
    15 EC + 35 EMC + 1 LiPF$_6$ & 2.6 $\pm$ 1.3 & 5.1 $\pm$ 1.1 \\
    \hline \hline
  \end{tabular}
  \caption{Calculated Li$^+$ diffusion coefficients  in each electrolyte, from the slope of the mean-square displacement (MSD) and integral of the velocity autocorrelation function (VACF). For each electrolyte, the most stable solvation configuration(s) were considered.
  }
  \label{tab:LiDiffusion}
\end{table}

Furthermore, analysis of the range of trajectories shows that
 slower diffusion tends to occur for cases where the Li$^+$ solvation structure is more energetically preferred.
A similar observation is noted when the coordination number is greater than 4 [\tit{i.e.,} the non-tetrahedral EMC case in Fig.~\ref{fig:EMCSolvation}(c)].
These results reveal that the magnitude of the diffusion coefficient is strongly dependent on how tightly solvated the Li$^+$ is by its solvent molecules.
When the Li$^+$ is more tightly solvated, the diffusion coefficient is smaller than when it is weakly solvated.
We  conclude that EC solvates Li$^+$  better than EMC, as indicated by the lower diffusion coefficient.

Table~\ref{tab:PDiffusion} shows the diffusion coefficients for PF$_6^-$, calculated by tracking the P atom.
Overall, the values are larger than for Li$^+$, consistent with the weaker solvation structure discussed above.
We also note the higher diffusivity in EMC compared to EC, which is due to the weaker solvation by EMC as evidenced by the shorter first-shell solvent molecule residence time.

\begin{table}
  \vspace{11pt}
  \centering
  \begin{tabular}{c c c c c}
    \hline \hline
   & \multicolumn{2}{c}{PF$_6^-$ Diffusion Coefficient}  \\
   & \multicolumn{2}{c}{(10$^{-6}$~cm$^2$/s)}  \\
  \cline{2-3}
      Electrolyte composition & MSD & VACF \\
    \hline
    63 EC + 1 LiPF$_6$ & 7.1 $\pm$ 0.9 & 9.2 $\pm$ 1.0 \\
    42 EMC + 1 LiPF$_6$ & 30.8 $\pm$ 8.8 & 28.6 $\pm$ 5.7 \\
    15 EC + 35 EMC + 1 LiPF$_6$ & 5.7 $\pm$ 2.4 & 9.5 $\pm$ 1.4 \\
    \hline \hline
  \end{tabular}
  \caption{Calculated PF$_6^-$ diffusion coefficients in each electrolyte, from the slope of the mean-square displacement (MSD) and integral of the velocity autocorrelation function (VACF). For each electrolyte, the most stable solvation configuration(s) were considered.
  }
  \label{tab:PDiffusion}
\end{table}

\subsection{Finite Size and Time Scale Effects}

Molecular dynamics simulations based on traditional Kohn-Sham density functional theory implementations are limited to moderate  system sizes on the order of hundreds of atoms and time scales of 10s of picoseconds.
In order to gauge finite size and time scale effects on the solvation structures and diffusion coefficients that we calculated using DFT, 
we used the \reax{} reactive force field\cite{reaxff, IslBryVan, BedSmiVan} as implemented in the LAMMPS\cite{lammps, reax_lammps} software package to run much longer molecular dynamics trajectories of 1~ns with system sizes up to $\sim$6400 atoms.

First, we assessed the quality of the \reax{} force field to reproduce the results of DFT, using the EC electrolyte as a test case.
These simulations were performed under NVT conditions using a Nos\'e-Hoover chain thermostat with 3 Nos\'e-Hoover chains and time step of 0.25~fs.
The Nos\'e frequency was set to $\sim$1333~cm$^{-1}$ corresponding to a period of $\sim$25~fs.
A system of 630 EC and 10 LiPF$_6$ molecules was equilibrated for 125~ps at 330~K followed by 1~ns of simulation time.
This test system contained the same concentration of LiPF$_6$ as the DFT simulation, but with 10 times more Li ions to gather better statistics.

\begin{figure}
   \centering
   \includegraphics[width=1.0\textwidth,clip=true]{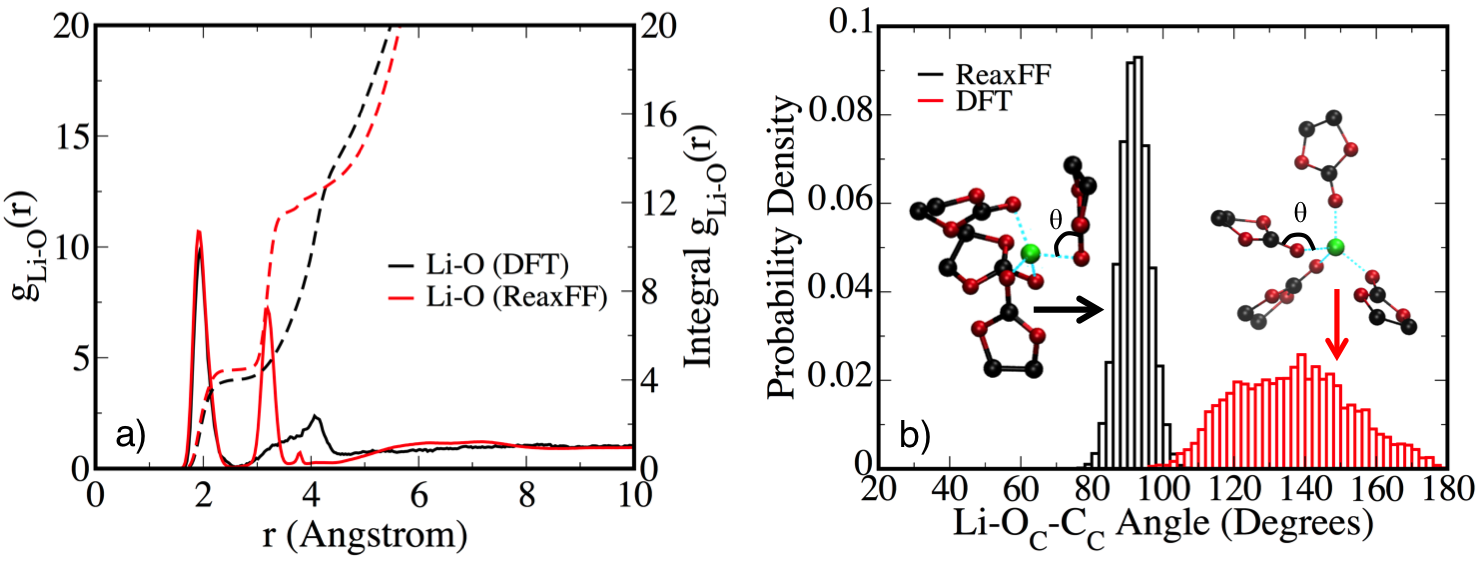}
   \caption{(a) Li--O pair correlation functions (solid lines) and their integrals (dashed lines) for dissociated LiPF$_6$ in EC, comparing DFT and \reax{} trajectories. (b) Histogram indicating the distribution of the Li--O$_C$--C$_C$ angle for \reax{} and DFT trajectories. Insets show representative snapshots of the \reax{} and DFT Li$^+$ solvation structures.}
   \label{fig:DFTReaxFF}
\end{figure}

In Fig.~\ref{fig:DFTReaxFF}(a), we show the Li--O pair correlation function and its integral for both the \reax{} trajectory and the DFT trajectory. 
We find that both methods predict the same coordination number of $\sim$4 carbonyl oxygen atoms  for the first solvation shell.
Also, the O$_C$--Li--O$_C$ angle distributions are  nearly identical, with a peak at $\sim$110$^\circ$ indicating a tetrahedral arrangement, as shown for example in Fig.~\ref{fig:ECOrientation}.
However, the second peak in the Li--O pair correlation function (associated with the ether oxygen atoms) is slightly different, with DFT predicting a broader peak centered further away than \reax{}.
This second peak is much sharper and more structured with \reax{}.
We can understand this difference more deeply by examining the Li--O$_C$--C$_C$ angle. 
Figure~\ref{fig:DFTReaxFF}(b) shows
that DFT exhibits  a broad distribution of angles centered at $\sim$140$^{\circ}$, while \reax{} predicts a narrow distribution of angles around $\sim$90$^{\circ}$.
Representative snapshots of these solvation structures from \reax{} and DFT are displayed in the insets of Fig.~\ref{fig:DFTReaxFF}(b),
which illustrate how the 90$^\circ$ angle from \reax{} results in the ether oxygen atoms being closer to Li$^+$ than in the DFT simulations.
In addition, the second Li--O peak in the pair correlation function from \reax{} is much sharper than from DFT,
because this ``bent'' solvation structure is more rigid, presumably from additional interactions between Li and O$_E$.
In the DFT simulations, the carbonyl dipoles of the EC molecules point toward the Li$^+$, but also exhibit more rotational fluctuations, giving rise to the broadening of the second peak in the Li--O pair correlation function and of the Li--O$_C$--C$_C$ angular distribution.

\reax{} also predicts occasional EC dimerization over the course of the trajectory where the carbonyl carbon of one EC molecule interacts with the carbonyl oxygen of another EC molecule.
This dimerization is inconsistent with DFT and the chemical inertness of the EC liquid, but only occurs for less than 5\% of the molecules over 1~ns.
We further compared the diffusion coefficients for both Li$^+$ and PF$_6^-$ 
and found that the \reax{} values are within $\sim$40--50$\%$ of the DFT values.
Thus, we do note differences between DFT and \reax{} for these systems, but the applicability of \reax{} to study finite size and time effects appears valid. 

\begin{figure}
   \centering
   \includegraphics[width=1.0\textwidth,clip=true]{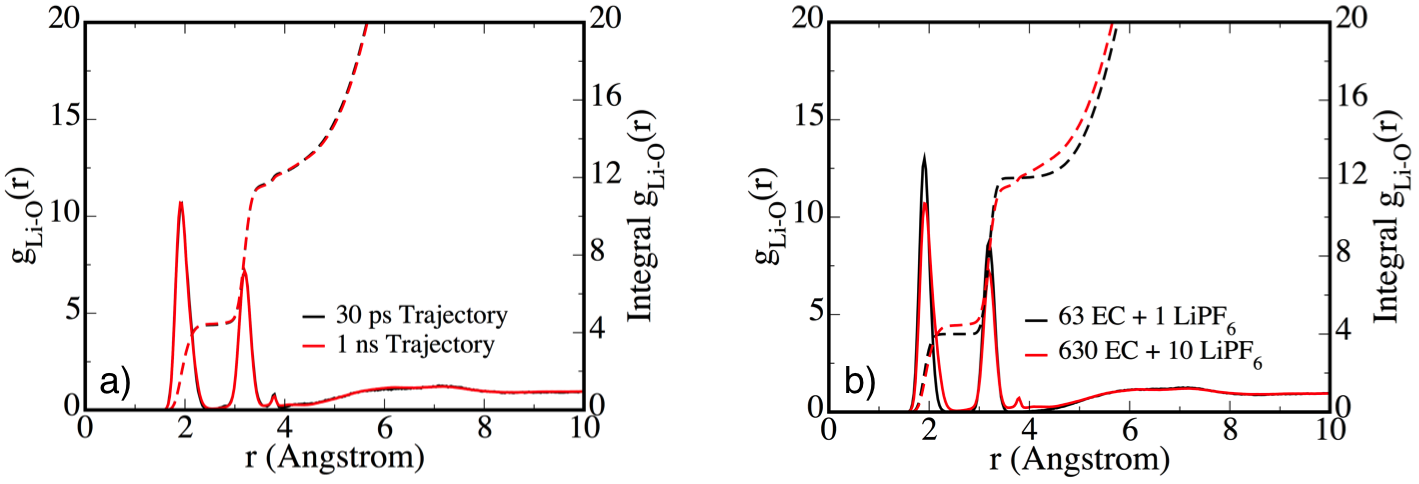}
   \caption{(a) Comparison of Li--O pair correlation functions and their integrals for 30 ps trajectory and 1 ns trajectory. (b) Comparison of Li--O pair correlation functions and their integrals for a 63 EC + 1 LiPF$_6$ system size and 630 EC + 10 LiPF$_6$ system size, where the concentration of the two systems are fixed.}
   \label{fig:time_size}
\end{figure}

To determine the effect of time scale on solvation structure, we calculated the Li--O pair correlation function and its integral using a small 30 ps segment of the 630 EC + 10 LiPF$_6$ 1~ns \reax{} trajectory.
In Fig.~\ref{fig:time_size}(a), these results are compared to the pair correlation function obtained when the entire 1 ns trajectory is used. 
We see that peak locations, intensities, and coordination number are very similar, indicating that the time scale used for the MD simulations does not have a significant effect on the solvation structure.
Similarly, to determine the effect of finite size on the solvation structure, we ran a 63 EC + 1 LiPF$_6$ MD simulation with \reax{} for 1 ns under NVT conditions at 330K.
The Li--O pair correlation function and integral for this 63 EC + 1 LiPF$_6$ system are compared to that calculated for the 630 EC + 10 LiPF$_6$ system in Fig.~\ref{fig:time_size}(b).
We see that although the peak intensities vary slightly, the peak locations and the integral of the pair correlation function are comparable.
These results were confirmed with several additional system sizes as well.
Based on these observations, we don't expect the solvation structures calculated using DFT to change significantly when going to longer time scales or larger system sizes at fixed concentrations.

We also examined the effects of time scale and finite size on the diffusion coefficients for both Li$^+$ and PF$_6^-$ in pure EC using \reax{}.
We compared the diffusion coefficient for Li$^+$ and PF$_6^-$, calculated using the slope of the mean square displacement, from a 30~ps segment of the 630 EC + 10 LiPF$_6$ trajectory to that from the entire 1~ns trajectory.
We found that using only 30~ps of the trajectory resulted in a difference of $\sim$46$\%$ for Li$^+$ and 38\% for PF$_6^-$ as opposed to using the whole 1 ns trajectory in the calculation.
Based on these results, we expect that our DFT-calculated diffusion coefficients may vary up to 50$\%$  by running longer simulations.
Classical simulations have also been used previously to study size effects, which found a 10$\%$ difference in the diffusion coefficient between large (480 solvent molecules) and small (240 solvent molecules) box sizes.\cite{BorSmi09} 
However, more Li atoms were included in these simulations, which increased the statistical sampling and resulted in correspondingly smaller finite-size effects.
We further compared the diffusion coefficients calculated from 1~ns trajectories of systems containing either 63 EC + 1 LiPF$_6$ or 630 EC + 10 LiPF$_6$, to judge the effects of finite size without changing the statistical sampling. 
The diffusion coefficients between the smaller and larger systems differed by $\sim$6$\%$ for Li$^+$, but up to $\sim$48$\%$ for PF$_6^-$.
Although the discrepancy for PF$_6^-$ is fairly large, it is comparable with the approximately 50$\%$ uncertainty we find for the time scale effects.
Tests on EMC and the EC/EMC mixture showed similar results as EC.

To summarize, we find that the Li$^+$ solvation structures do not change significantly when assessing finite size and time scale effects, although there are some differences between the \reax{} and DFT solvation structures.
Furthermore, uncertainties up to $\sim$50\% in the diffusion coefficients are expected for simulations run at short time scales and smaller length scales.
This uncertainty is still small enough to allow us to draw qualitative and semi-quantitative conclusions as above, given the relative statistical errors in our computed values.
However, it will be important in the future to use DFT at still larger length and time scales to further reduce uncertainties in solvation structures and diffusion coefficients.

\section{Conclusions}

We found multiple possible solvation structures of Li$^+$ in each of the electrolytes studied here, including ethyl carbonate (EC), ethyl methyl carbonate (EMC), and EC/EMC mixture. 
While previous literature on EC and LiPF$_6$ has focused on solvation by 4 carbonyl oxygen atoms, we observed solvation structures that also include PF$_6^-$ in the first solvation shell. 
We found that the preferred solvation structure is strongly dependent on the solvent.
For EC, Li$^+$ prefers to be solvated by 4 carbonyl oxygen atoms and to remain dissociated from PF$_6^-$, while for EMC it shows some preference for only 3 carbonyl oxygen atoms and to remain close to the PF$_6^-$.
The 3:7 EC/EMC mixture shows a preference for Li$^+$ to remain dissociated from PF$_6^-$, but has two energetically similar solvation structures where Li$^+$ is solvated either by 4 carbonyl oxygen atoms or by 3 carbonyl oxygen atoms and 1 ether oxygen atom. 
In all cases, Li$^+$ prefers to be solvated in a tetrahedral arrangement, though this does not rule out the possibility of a slightly higher-energy non-tetrahedral structure forming, as seen in the EMC ``4etherPF6'' case. 
Comparisons of solvation structures for Li$^+$ and PF$_6^-$ reveal that Li$^+$ is more strongly solvated, associated with lower mobility in the electrolyte.
Calculations of first shell solvent molecule residence times show that there is a slight preference for PF$_6^-$ to be solvated by EC over EMC, although both show weak solvation of PF$_6^-$.

Calculated diffusion coefficients quantify the ionic motion in each electrolyte and relate to the solvation structures.
We find that the largest Li$^+$ diffusion coefficient occurs in EMC. 
This is consistent with the measured viscosities of both organic solvents, and the diffusivity values obtained agree well with experimental values. 
The magnitude of the diffusion coefficient is largely influenced by how tightly the ion is solvated, for both Li$^+$ and PF$_6^-$.
A more tightly bound solvation structure, such as for Li$^+$ in EC, leads to slower diffusion of the solvated ion.
Comparison of Li$^+$ and PF$_6^-$ diffusion coefficients indicate that PF$_6^-$ diffuses faster than Li$^+$ in all the electrolytes examined here, even though it is the heavier species.
This can be attributed to the fact that the solvent molecules interact more strongly with Li$^+$ than with PF$_6^-$.

Furthermore, we quantified finite size and time scale effects  on the solvation structures and diffusion coefficients using \reax{}.
We find that solvation structures do not change significantly for larger system sizes and longer time scales than used here with DFT, but there are some structural differences between \reax{} and DFT simulations.
Absolute diffusion coefficients are more affected by size and time scale effects, 
as uncertainties can be as large as 50\%, but relative values remain consistent.

Our work has shown that a more tightly bound solvation structure leads to slower diffusion, and a weakly bound solvation structure leads to faster diffusion. To improve the mobility of Li ions in solution, our results suggest that Li$^+$ must have weak interactions with the organic solvent used in the electrolyte. This is valuable insight that can be used to improve the cycling rate of Li-ion batteries and potentially lead to the design of new electrolytes for better overall battery performance.

\begin{acknowledgement}

MTO would like to acknowledge Brandon Wood, Kyle Caspersen, Eric Schwegler, Tingting Qi, and Md Mahbubul Islam for useful comments and discussions regarding content presented in this manuscript.
This work was performed under the auspices of the U.S. Department of Energy by Lawrence Livermore National Laboratory under Contract DE-AC52-07NA27344. Support for this work was provided through Scientific Discovery through Advanced Computing (SciDAC) program funded by U.S. Department of Energy, Office of Science, Advanced Scientific Computing Research and Basic Energy Sciences.  
OV and ACTvD acknowledge funding from a grant from the U.S. Army Research Laboratory through the Collaborative Research Alliance (CRA) for Multi Scale Multidisciplinary Modeling of Electronic Materials (MSME) and from  the Fluid Interface Reactions, Structures and Transport (FIRST) Center, an Energy Frontier Research Center funded by the U.S. Department of Energy, Office of Science, and Office of Basic Energy Sciences.

\end{acknowledgement}


\input{Ong_LiBatt_Rev.bbl}

\end{document}

%% file: Ong_LiBatt_Rev.bbl
\providecommand*\mcitethebibliography{\thebibliography}
\csname @ifundefined\endcsname{endmcitethebibliography}
  {\let\endmcitethebibliography\endthebibliography}{}